\documentclass[conference]{IEEEtran}
\IEEEoverridecommandlockouts

\usepackage{cite}
\usepackage{amsmath,amssymb,amsfonts}
\usepackage{algorithmic}
\usepackage{graphicx}
\usepackage{textcomp}
\usepackage{tabularx}
\usepackage{url}
\usepackage{xcolor}
\usepackage{booktabs}
\usepackage{multirow}
\usepackage[hidelinks]{hyperref}

\usepackage{tikz}
\usetikzlibrary{arrows.meta,positioning,shapes.geometric,shapes.misc,fit}

\usepackage{listings}
\usepackage{xcolor}
\definecolor{dkgreen}{rgb}{0,0.5,0}
\definecolor{gray}{rgb}{0.5,0.5,0.5}
\definecolor{mauve}{rgb}{0.58,0,0.82}
\usepackage[most]{tcolorbox} 
\tcbset{
  resbox/.style={
    colback=gray!5,
    colframe=gray!60!black,
    boxrule=0.5pt,
    arc=2pt,
    left=6pt,
    right=6pt,
    top=6pt,
    bottom=6pt,
    enhanced
  }
}

\usepackage[ruled, linesnumbered]{algorithm2e}

\newcommand{\tool}{Prex}
\newcommand{\category}[1]{\textsc{\footnotesize{}#1}} 

\usepackage{xparse}
\ExplSyntaxOn
\NewDocumentCommand{\code}{m}
 {
  \group_begin:
  \footnotesize\sffamily
  \tl_set:Nn \l_tmpa_tl { #1 }
  \tl_replace_all:Nnn \l_tmpa_tl { < } { \textless{} }
  \tl_replace_all:Nnn \l_tmpa_tl { > } { \textgreater{} }
  \tl_replace_all:Nnn \l_tmpa_tl { | } { \textbar{} }
  \tl_use:N \l_tmpa_tl
  \group_end:
 }
\ExplSyntaxOff

\begin{document}

\title{Pattern-Based Precondition Inference\\
from Method Implementations
}

\author{
  \IEEEauthorblockN{Shanyue Zhou and Martin P. Robillard}
\IEEEauthorblockA{
  \textit{School of Computer Science}\\
  \textit{McGill University}\\
  Montr\'{e}al, QC, Canada} \\
  ORCID: \href{http://orcid.org/0009-0004-3875-3275}{0009-0004-3875-3275}, \href{http://orcid.org/0000-0002-0248-1384}{0000-0002-0248-1384}
}

\maketitle

\begin{abstract}
Method preconditions define the conditions that must be satisfied before a method is called. Often, such conditions are not documented and must be inferred. We present \tool{}, a pattern-based static analysis approach that infers preconditions from explicit checks in Java source code. For each precondition, \tool{} reports the condition required for a valid call and the reported exception in the case of a violation. \tool{} also attempts code transformations to convert the condition detected within a method's implementation into an expression that can be evaluated by client code. It leverages the concept of a mutation boundary to separate precondition checks from exceptions that may occur after the method changes the program state. We evaluated \tool{} from 46 open source Java projects and manually reviewed independent samples of 150 reports. Among the 150 \tool{} reports, 149 were correct and 125 of the correct reports were executable by client code. In a systematic comparison with an alternative approach based on path expansion, \tool{} provided more precise results. In total, 122 of the 149 correct reports had no relevant method documentation, and 41 required nontrivial source code analysis to recover manually. Overall, explicit source-code checks provide precondition information that can be absent from method documentation, and \tool{} can help recover it.
\end{abstract}


\section{Introduction}

Developers routinely define preconditions on the execution of methods. Preconditions are an important enabler of software understandability and quality as they provide a precise definition of the input that is valid to provide a method~\cite{161279, 7194623}. 
In Java, developers can indicate preconditions through documentation, source code and, in a limited sense, annotations~\cite{Monperrus_Eichberg_Tekes_Mezini_2012}. However, in many software projects, there is no formal support for documenting and reasoning about preconditions, and usage is haphazard~\cite{6200082}.

In this context, users of software packages are often left uncertain about the preconditions for calling a method. If the documentation is lacking or incorrect, violated preconditions can lead to exceptions and undefined behaviors. Avoiding this eventuality can require defensive programming in user code~\cite{10.1145/1572272.1572284}. The goal of our research is to provide automated, precise, and deterministic support for identifying preconditions in existing source code. This paper contributes a solution for the Java language based on general principles intended to carry over to comparable programming languages.

We present \tool{} (Precondition Extractor), a tool-supported static analysis approach to extract explicit precondition checks from Java source code and convert them into \textit{precondition descriptors}. Each descriptor records the \textit{guard} condition required for a valid call, the violation \textit{signal}, and a mapping from guard symbols to their method-boundary representations. The analysis recognizes common Java precondition patterns, including assertions and conditional throws. \tool{} leverages the concept of a \textit{mutation boundary} to exclude runtime failures that can occur only after the method has started changing program state.
Finally, it attempts to express guards using method arguments and public query methods so that client code, which calls the analyzed method, can evaluate them before invocation. We call such guards \textit{testable}. Some preconditions depend on hidden state for which no public expression exists; \tool{} identifies such expressions as \textit{untestable}.

The design of \tool{} is conservative. If it cannot establish that a candidate is an explicit precondition check before the mutation boundary, it discards the candidate. Every reported descriptor can be traced to the source expressions and control-flow decisions used to construct it. This makes the analysis deterministic and allows its results to be checked and reproduced. The approach does not rely on generated text, training data, or runtime executions.

The closest alternative to the \tool{} approach is WIT, which uses bounded path analysis to extract conditions that lead Java methods to throw exceptions~\cite{MF2023a}. WIT targets exception behavior: a reported condition explains when an exception can be reached. Our goal is more specific, as we seek explicit checks that define obligations on the user before method execution. An exception path is therefore excluded if it crosses the mutation boundary, even if its path condition is valid. 
Other work mines preconditions from repeated guards at API call sites or infers broader program properties from executions~\cite{Nguyen_Dyer_Nguyen_Rajan_2014,Ernst_Perkins_Guo_McCamant_Pacheco_Tschantz_Xiao_2007}. In contrast, our evidence comes from explicit checks in the implementation of the target method and its helpers. This choice limits recovery to explicit preconditions, but it ties every precondition report to a specific check in the source code and does not require a usage corpus or test suite.

We evaluated \tool{} on methods from 46 open-source Java projects and manually reviewed two independent samples of 150 precondition descriptors: one sampled from \tool{} and one sampled from WIT. Of the 150 \tool{} reports, 149 (99.3\%) were correct. Among these correct reports, 125 were testable by users. This includes 12 cases in which the source check used hidden state but \tool{} recovered a public expression for the same state. In contrast, only 57 of the 150 sampled WIT reports were true and correctly-expressed preconditions. Most of the false positives were caused by exception paths that crossed the mutation boundary. Our results also show that precondition extraction is important to support developers, as we found that 122 of the 149 correct \tool{} reports identified preconditions that were not documented.

This paper makes the following contributions:
\begin{itemize}
    \item A novel static analysis algorithm for extracting preconditions from the computation of a mutation boundary and the detection of recognized patterns in source code;
    \item A technique to transform guard condition expressions to use method-boundary elements and to determine whether the resulting conditions are testable by client code;
    \item An empirical evaluation of the precision and testability of the output of the approach on 46 Java projects and a systematic comparison with a path expansion approach;
    \item Empirical data about the gap between implicitly checked and documented preconditions.
\end{itemize}

The rest of the paper is organized as follows. In Section~\ref{s:Relwork}, we review research on exception behavior and precondition mining and explains how our work differs. Section~\ref{s:Approach} formulates the problem and describes the design and implementation of \tool{}. Section~\ref{s:Evaluation} presents our evaluation dataset, manual review process, and results. Finally, in Section~\ref{s:Conclusion} we place the findings in perspective and discuss future work.

\paragraph*{Data availability} We released Prex as a free and open source software project~\cite{Prex} and the research data we collected is available on Zenodo~\cite{replicationPackage}.
\section{Related Work}
\label{s:Relwork}

Research on Java preconditions has followed several related directions. The direction closest to the present paper studies the static extraction of exception conditions from method implementations. Another direction mines likely preconditions from recurring guard patterns at API call sites across large Java code corpora. There is also work on code inference based on contracts, annotations, and invariants. Across these directions, the main distinction from our work is that we focus on explicit precondition checks in method implementations, rather than on all exception-triggering conditions or broadly-inferred contracts. 

\subsection{Exception Behavior Extraction}

Several prior works extract exception conditions by reconstructing paths that lead to thrown exceptions. Marcilio and Furia presented WIT, a tool that extracts the exception preconditions of Java methods and constructors. WIT statically analyzes the paths in a method's implementation that lead to an exception being thrown~\cite{MF2023a}. WIT is different in that it seeks to identify \textit{exception preconditions}, i.e., an input configuration that triggers an exception. To do so, WIT builds paths that end in explicit throws or exceptional calls, inlines callees when their source code is available, and checks whether the resulting path condition is feasible. It uses bounded path exploration to extract the full path of the condition without building it. For this purpose, it relies on an encoding of paths limited to a core subset of Java features and library
methods. 

WIT reports conditions that can lead a method to throw an exception. In contrast, we report a condition only when it acts as a precondition that users should satisfy before calling the method. For example, an exception may happen only after the method has already changed state, or the extracted formula may refer to internal state that users cannot directly test. We therefore use the mutation boundary and testability analysis to decide what and how to report as a precondition. In contrast to WIT, we aim to report only preconditions that users need to understand and validate before they call the method. 

Buse et al. developed a technique to generate exception documentation based on interprocedural exception propagation analysis with symbolic execution-like path reasoning~\cite{10.1145/1390630.1390664}. The main connection to our research is that it statically extracts exception-related paths and turns them into documentation, again without the focus on producing precondition descriptors testable by client code. 

Related work on software testing addresses exception conditions from a different perspective: rather than inferring conditions from code, it translates documented \code{@throws} conditions or specification templates into executable checks for tests. Even though that work is not a code-based inference technique, it shows that exception conditions and explicit predicates can be interchangeable~\cite{Goffi_Gorla_Ernst_Pezzè_2016, NHS2022a}.

Overall, work on exception behavior extraction can be placed in the broader context of exception flow and error-handling behavior analysis~\cite{10.1145/941566.941569, Acharya_Xie_2009, Weimer_Necula_2005, 10.1007/3-540-48166-4_20}. This work investigates exception-triggering paths, propagation, and error-handling behavior in method implementations.

\subsection{Precondition Mining from Client-Side Patterns}

Another line of work extracts preconditions and constraints from recurring guard patterns at API call sites.

Prior work infers API preconditions by mining guard conditions in client code across large Java corpora~\cite{10.1145/1273442.1250749}. The basic idea is that if many call sites use the same guard before a call, that condition may reflect a precondition of the target API~\cite{Nguyen_Dyer_Nguyen_Rajan_2014}. Follow-up work further develops this consensus-based view that recurring client-side guards can reveal likely API preconditions~\cite{Nguyen_Dyer_Nguyen_Rajan_2015, 10.1145/3278177.3278182}.

Other empirical work studies parameter constraints in Java libraries and where they appear, such as in Javadoc or implementation code~\cite{Zhong_Meng_Li_Jia_2020, 10.1145/2803171}. The parameter rules are similar to preconditions, but may not have explicit guards in the code. In this case, the evidence comes from recurring client-side guard patterns, and the recovered conditions are inferred from usage rather than from the target method itself. In contrast, we recover explicit checks directly from method implementations, which makes the result deterministic even when usage data is limited.

\subsection{Contract, Annotation, and Invariant Inference}

Beckman et al. developed ProtocolFinder, a tool that identifies code locations that may indicate object protocols~\cite{BKA2011a}. ProtocolFinder is related to our work because it also uses source code patterns as evidence of usage constraints in Java programs. ProtocolFinder reports object protocols, which specify the temporal ordering of method calls, while \tool{} reports preconditions that must hold before each individual method call.

Flanagan et al. introduced Houdini, which infers candidate annotations with the modular static checker ESC/Java~\cite{Flanagan_Leino_2001, 10.1145/543552.512558}. Houdini generates a large set of candidate annotations, then iteratively runs ESC/Java to refute and remove candidates until a fixed point is reached. Even though Houdini is not focused on preconditions, it provides a scalable way to run template-like filters to infer annotations.

It is also possible to infer invariants through dynamic inference systems such as Daikon~\cite{Ernst_Perkins_Guo_McCamant_Pacheco_Tschantz_Xiao_2007, 10.1145/302405.302467}. These systems infer likely invariants, including procedure preconditions, from observed executions. They show that precondition-like properties can be recovered from program behavior, but they depend on execution traces rather than on explicit checks in code. Numerous other works infer specifications based on code and documentation, but they all are less directly related to our focus on explicit precondition extraction triggered by exceptions~\cite{10.1007/978-3-642-18275-4_12, Flanagan_Joshi_Leino_2001, Zhou_Wang_Yan_Chen_Panichella_Gall_2020, 10.1007/978-3-642-35873-9_10}. These works extract evidence from inferred protocols, annotations, or execution traces, and the goal is to recover broader contracts rather than explicit method-entry checks.

\section{Approach}
\label{s:Approach}

We developed \tool{}, a source code analysis tool to detect preconditions from the source code of Java methods and produce guard expressions testable by client code.

\subsection{Scope and Terminology}

Generally speaking, a \textbf{precondition} is any predicate that must be true before a method is executed. In principle, executing a method without respecting all of its preconditions leads to undefined behavior. We refer to the execution of a method with one or more precondition violations as an \textbf{illegal execution}. In practice, the result of an illegal execution is often a runtime error indicated through Java's exception-handling mechanism. Other possible results include silently incorrect behavior and system crashes.

In the implementation of a method, a precondition can be either checked or unchecked. This distinction does not relate to the type of exception raised, but rather to whether there is code explicitly checking that the precondition holds. A precondition is checked if the method's implementation includes an explicit validation of input arguments or object state that aborts the method execution upon failure. In Java this check is typically implemented via a \code{throw} or \code{assert} statement. For example, the first statement of many of the implementations of \code{List\#getFirst} check whether the list is empty and explicitly throw an exception if it is. We refer to such code fragments as \textbf{precondition checks}. We define a precondition check as a predicate on the input to a method that occurs before the method has any side effect and that results in the execution of the method becoming \textit{disabled}. Here, a side effect is a change to state outside the method's local variables\cite{10.1007/978-3-540-30579-8_14, 10.1145/3238147.3238226}. This includes updates to object fields. We call the \textbf{mutation boundary} the first reachable control-flow node on each path from the method entry point that is estimated to correspond to code with a side effect. For direct state mutations, this node is the first write to an instance or static field. A precondition check comprises two parts: a \textbf{guard}, which must hold for a valid call, and a violation \textbf{signal}, which is the type of exception raised when the guard does not hold. A precondition can also be unchecked in the sense that, although it is documented and/or its violation results in an exception being thrown, it is not possible to unambiguously identify a dedicated check for it.

The guards in precondition checks can be further categorized as \textbf{testable guards} or \textbf{untestable guards}. A testable guard can be evaluated by client code before calling the method. An untestable guard depends on information that is inaccessible outside the class.

Finally, preconditions can be documented with various degrees of quality. \textbf{Precise documentation} includes both a testable guard and signal. \textbf{Partial documentation} includes only fragments of information or untestable code. In many cases, only a \code{@throws} clause is provided in the Javadocs. Finally, preconditions can, of course, remain \textbf{undocumented}.

Given the above terminology, we define the scope of our approach as a tool aimed at converting precondition checks into precise documentation, i.e., Javadocs that include testable guards and the violation signal.

\subsection{Problem Formulation}

We define the problem of precondition extraction as a function that takes as input a Java method and produces a set of \textbf{precondition descriptors}. We model a precondition descriptor $p$ as a tuple $(g, s, V)$, where:

\begin{itemize}
    \item $g$ is the precondition guard, i.e., the boolean predicate that must be true to avoid invalidating the method's execution. In documentation, the guard is typically negated to describe the condition under which the violation signal occurs.
    \item $s$ is the violation signal, expressed as the fully-qualified type of the exception (or error) thrown if the guard evaluates to false in an execution of the method. This element refers to a subtype of \code{Throwable}. 
    \item $V$ records how client code can obtain the values used in $g$ before the call. A parameter is marked DIRECT because its value is directly accessible by the method caller. Other symbols map to equivalent public field accesses or public query expressions when available, and to HIDDEN when no accessible equivalent is found.
\end{itemize}

Listing~\ref{lst:precondition_output} shows the precondition descriptor extracted for method \code{ArrayList\#add} of the Java Development Kit (JDK, version 25). The field \code{input} identifies the input method, and the \code{guard}, \code{signal}, and \code{map} correspond to elements $g$, $s$, $V$, respectively. The map shows that the public getter method \code{size()} has been identified for symbol \code{size} and that symbol \code{index} refers to a parameter. The field \code{description} combines the elements into a precise piece of documentation. 

\begin{lstlisting}[caption={Example precondition output produced by \tool{}. Package names of Java elements are omitted for readability}, label={lst:precondition_output}]
  ``input'' : ``ArrayList#add(index, E)'',
  "guard" : "!(index > size || index < 0)",
  "signal" : "IndexOutOfBoundsException",
  "map" : {
    "size" : "size()",
    "index" : "DIRECT"
  },
  "description" : "@Throws IndexOutOfBoundsException if index > size() || index < 0",
  "testability" : "TESTABLE",
\end{lstlisting}

\subsection{Analysis Algorithm}
\label{ss:Algorithm}

\tool{} extracts precondition descriptors from a collection of Java source files. In principle, there is no requirement for the input files to logically correspond to a single project. However, the input collection is considered as a unit for the purpose of resolving dependencies, as described below. Henceforth, we refer to the input collection of files simply as a \textit{Java program}.

The analysis begins by parsing the program source code into a collection of abstract syntax tree (ASTs) using the JavaParser library.\footnote{\url{https://javaparser.org/}. We used version 3.27.1.} Algorithm~\ref{alg:mainStrategy} presents the overall strategy to extract precondition descriptors from the input program. In addition to a Java program, the analysis takes as input a customizable catalog of common precondition-checking methods that are not part of the input Java program (e.g., \code{Objects\#requireNonNull}).

\begin{algorithm}[ht]
\caption{Overall Analysis Strategy}
\label{alg:mainStrategy}
\KwIn{${J}$: Java program as ASTs}
\KwIn{${L}$: Precondition catalog}
\KwOut{$P$: Set of all precondition descriptors for $J$}

$T \gets \mathrm{extractTypeHierarchy}(J)$\;
$P \gets \emptyset$\;

\ForEach{method ${m} \in {J}$}{
    $C \gets \emptyset$\tcp*{Potential Preconditions}
    $P_m \gets \emptyset$\;
    $C \gets \mathrm{preconditionCandidates}(m, T, L)$\;
    $C \gets \mathrm{removeNonPreconditionFailures}(C)$\;
}

\ForEach{candidate $c \in C$}{
        $c' \gets \mathrm{propagateDownstreamChecks}(c, C, {L})$\;
        \If{$\mathrm{isPublic(declaringMethod}(c'))$}{
            $s \gets \mathrm{extractSignal}(c')$\;
            $g \gets \mathrm{extractGuard}(c')$\;
            $V \gets \mathrm{resolveVariableAccess}(c', {T})$\;
            $p \gets (g, s, V)$\;
            $P_m \gets P_m \cup \{p\}$\;
        }
    }
    $P \gets P \cup P_m$\;
\end{algorithm}

Before processing individual methods, the \code{extractTypeHierarchy} function performs a preliminary pass over the ASTs of all compilation units to create a data structure that comprises all classes, their fields and visibility, and inheritance relations between classes (Line~1). This global context information is required to accurately resolve whether instance fields are accessible to client code.

The analysis then proceeds in a modular fashion by inspecting each method individually (Line~4). \tool{} first identifies potential precondition candidates in the method (Line~6). At this stage, a \textit{precondition candidate} is either a recognized code pattern that results in an exception, or any method call (which could result in a downstream precondition check). The function \code{removeNonPreconditionFailures} filters these candidates using the estimated mutation boundary of each method and removes all non-precondition checks that lie beyond the boundary (Line~7).

For each candidate, we then inline any precondition check from downstream methods into the calling method's context, ensuring that the precondition is attributed to the method that exposes it. As part of this propagation, the algorithm performs argument substitution so that the downstream method's local conditions are accurately expressed in terms of the calling method's scope. Once downstream methods are inlined, the candidate is considered a complete precondition check ($c'$, Line~10).

In the final stage of the algorithm, we check if the precondition check's declaring method is public and, in the affirmative, we convert our internal representation of a precondition check into a precondition descriptor (Lines~12-15) and add it to the output (Lines~16 and 19).

\subsection{Implementation Details}

\tool{} relies on several dedicated strategies to transform AST nodes into precondition descriptors and classify their testability. 

The precondition catalog ($L$) has two roles: identifying API methods that perform checks, such as \code{Objects\#requireNonNull}, and identifying public state-querying methods used in guards, such as \code{List\#size}. Every method listed in the catalog is assumed to have been manually verified as non-mutating. \tool{} includes selected methods from \code{java.util} and \code{java.lang} and allows users to extend the list with any method, including from project-specific code.

\subsection*{Recognized Patterns}

In \code{preconditionCandidates}, we use three patterns to cover all explicit precondition checks that throw exceptions using \code{throw} or \code{assert} statements. A precondition is either alway violated, or detected through a conditional check or an assertion statement. We convert the precondition checking process into three patterns: (1) unconditional throw, (2) assertion-based check, and (3) conditional throw. \tool{} is capable of handling variations in expressions of these general patterns, including early returns and preconditions raised in downstream method calls.

\subsubsection*{Unconditional Throw}

This pattern corresponds to a \code{throw} statement in the direct path of the method entry that raises an exception whenever the method is executed. This structure is typically used in abstract classes, or for unsupported interface methods.
For example, in Listing \ref{lst:case1}, the method \code{set} from \code{AbstractList} always throws an \code{UnsupportedOperationException}. This pattern corresponds to a precondition that can never be met.

\begin{lstlisting}[caption={Unconditional \code{throw} in \code{AbstractList}}, label={lst:case1}]
    public E set(int index, E element) {
        throw new UnsupportedOperationException();
    }
\end{lstlisting}

\subsubsection*{Assert-based Checks}

The second recognized pattern is the assertion statement. Each assert statement evaluates a boolean expression and throws an \code{AssertionError} if the condition is not satisfied (when assertion checking is enabled).

\subsubsection*{Conditional Throw}

This more general form of precondition check uses an \code{if-throw} pattern. Listing~\ref{lst:gen0} shows an abstract representation of this pattern. The \code{preGuardExecution()} and \code{postGuardExecution()} are respectively the code that executes before and after the precondition check. The expression in the \code{if} statement is the \code{throw} condition. \tool{} negates this condition when needed to produce the guard $g$ required for a valid call. The mutation boundary of \code{preGuardExecution()} will be discussed in the next section.

\begin{lstlisting}[caption={Abstraction of conditional throw pattern}, label={lst:gen0}]
    methodName(inputArgs) {
        preGuardExecution();
        if (throwCondition) {
            throw Exception e;
        }
        postGuardExecution();
    }
\end{lstlisting}

If the pre-guard execution does not cross the mutation boundary, we consider the \code{if-throw} block to represent a precondition. \tool{} then attempts to inline the elements used in the guard. The inlining logic aims to report the most accessible method or field that can be used to test the precondition. The depth of inlining is determined by the accessibility of the elements involved. For example, in Listing~\ref{lst:case3}, \code{len} uses the value of \code{size()}. Since \code{size()} is a public method without state mutation, the tool replaces the access to \code{len} with a call to \code{size()} and negates the \code{throw} condition to recover the testable guard \code{size() != 0}. In contrast, in Listing~\ref{lst:extract_testable}, the \code{throw} condition contains a private method \code{computeJ} that is not accessible, so the tool inlines the helper methods logic to recover a testable guard.

\begin{lstlisting}[caption={Conditional throw pattern precondition from class Stack}, label={lst:case3}]
    public synchronized E peek() {
        int len = size();
        if (len == 0)
            throw new EmptyStackException();
        return elementAt(len - 1);
    }
\end{lstlisting}

\begin{lstlisting}[caption={Input validation}, label={lst:extract_testable}]
    public void getResult(int i) {
        if (computeJ(i) > 10) {
            throw new IllegalArgumentException();
        }
    }

    private int computeJ(int num) {
        return num + 5;
    }
\end{lstlisting}

\tool{} supports variations in expressions of the conditional. In Listing~\ref{lst:case4}, the exception is placed in the \code{else} block. To extract the precondition, \code{preconditionCandidates} traverses the \code{if-else} branches upward from the \code{throw} statement, logically negating conditions to recover the required state. In this example, the extracted throw condition is
\begin{center}
    \code{!(initialCapacity > 0) \&\& initialCapacity != 0}
\end{center}
which the tool simplifies and negates to produce the guard \code{initialCapacity >= 0}. 

\begin{lstlisting}[caption={Cases where throw statement in the else block}, label={lst:case4}]
    public ArrayList(int initialCapacity) {
        if (initialCapacity > 0) {
            this.elementData = new Object[initialCapacity];
        } else if (initialCapacity == 0) {
            this.elementData = EMPTY_ELEMENTDATA;
        } else {
            throw new IllegalArgumentException("Illegal Capacity: "+
                                               initialCapacity);
        }
    }
\end{lstlisting}

The same traversal is applied to nested if statements and early returns. Starting from the \code{throw} statement, \tool{} walks through the control flow and collects the conditions that must hold to reach the violation signal. When a condition must be false to reach the signal, \tool{} negates the corresponding condition before adding it to the recovered guard.

\subsection*{Mutation Boundary}

A precondition must be evaluated before a method performs any externally observable state mutation. We define the \textit{mutation boundary} as the first reachable control-flow node on each path from the method entry point that is estimated to correspond to code with a side effect. 

Due to aliasing, abstraction, and polymorphism, it is practically impossible to determine \textit{precisely} whether a statement has a side-effect or not. To ensure that our approach remains conservative, our algorithm (\code{removeNonPreconditionFailures}) assumes that any statement has, by default, a side effect, and reverses this decision only if it can be verified that the statement has no side effect. To verify whether a statement is side-effect free, our analysis separately considers \textit{mutating operators} and \textit{method calls expressions}. 

If a statement involves one or more mutating operators, such as assignment, increment, or decrement, \tool{} checks the target of the operation. Mutation operations applied to method-local variables are allowed, but all others are considered to be mutating. For non-primitive local references, the analysis also tracks name aliases introduced before the check, such as a local variable assigned from a field. If the alias is later used to update the referenced object, the update is also considered crossing the mutation boundary.

For statements involving method calls, \tool{} checks whether each call can be treated as non-mutating before the guard. Calls in the precondition catalog are considered non-mutating by definition, as they must be manually validated as such. Calls to methods defined outside the analyzed program and outside the precondition catalog are conservatively considered to be mutating. This approach also covers unresolved calls that may perform file or network I/O, which the analysis does not separately model those resources. For calls to methods within the analyzed program, we first determine whether they are statically bound or not (i.e., calls to static, final, or private methods). If a method is dynamically dispatched, we conservatively consider it to be mutating. If the method is statically dispatched, we recursively traverse its body to determined whether we can consider it non-mutating using the same procedure as described above. Finally, as a special case, if a call expression is to a method of a non-final class defined in the analyzed program \textit{and} that has a single implementation, we analyze the single implementation instead of considering the method non-mutating, and issue a warning in the tool's output. During the development of the approach, we determined that excluding all dynamically dispatched calls was too conservative with respect to the mostly theoretical possibility that an external class could be injected by a third party.

\subsection*{Resolving Variable Access}

The \code{resolveVariableAccess} function evaluates the extracted guard to construct the variable mapping $V$ and determine the precondition's testability. The function iterates through every variable and method in the guard. If a variable corresponds to a method input argument, it is mapped as \code{DIRECT}. Fields that are public or exposed via a getter method are mapped to their respective representations, while internal fields lacking public accessibility are flagged as \code{HIDDEN}. A method is considered a getter method for a field only if it is public, takes no parameters, does not cross the mutation boundary, and directly returns a field from the same class. For example, the guard for \code{ArrayList\#getFirst} refers to the private field in \code{size != 0}. Mapping \code{size} to the getter method \code{size()} produces the testable guard \code{size() != 0}. The default implementation in \code{List\#getFirst} already uses the public expression \code{!this.isEmpty()}, so its guard is directly testable.

\subsection{Limitations and Exclusions}

There are hypothetical cases of precondition checks that \tool{} intentionally excludes from analysis. Silent failures are excluded, as the method does not throw an exception when a precondition is violated, but instead silently fails or produces undefined behavior. It is difficult to statically detect silent failures as there is no clear indication of failure. We also ignore exceptions handled in catch blocks, since they represent internal error handling rather than checks before execution. We also exclude predicates evaluated in loops because of the difficulty of precisely identifying the exact conditions under which the precondition holds using only static analysis.

Besides these intentional exclusions, the tool has a few known limitations. The \code{extractTypeHierarchy} function tracks arrays and collections as a whole, without tracking which element or nested object is mutated. 
This may make guards that depend on such state harder to recover. The static handling of downstream calls may miss checks that depend on dynamic dispatch or external code. Lastly, our detection of the mutation boundary can be imprecise when the method contains complex loops. To prevent false positives, we treat all loops before the precondition check as potentially mutating. Read-only loops may therefore cause valid precondition checks to be discarded.

\section{Evaluation}
\label{s:Evaluation}

We evaluated \tool{} through a benchmarking study and systematic comparison with a path-expansion approach. We analyzed precondition descriptors extracted from methods in 46 Java projects from the WIT dataset. The goal of our study was to provide a quantitative assessment of the estimated operational performance of pattern-based precondition extraction. We investigated the following questions:

\begin{itemize}
    \item \textbf{RQ1 (Precision)}: What proportion of preconditions reported by \tool{} are correct?
    \item \textbf{RQ2 (Testability)}: What proportion of correct preconditions reported by \tool{} is testable by client code?
    \item \textbf{RQ3 (Comparison)}: How does a pattern-based approach compare to a systematic path expansion approach?
    \item \textbf{RQ4 (Value)}: How does pattern-based precondition extraction complement existing documentation?
\end{itemize}

\subsection{Sampling Approach and Dataset}

We draw conclusions from the application of \tool{} to a sample of Java methods. We had three main requirements for our sampling approach. First, for the \textit{ecological validity} of our results, methods in the sample should constitute examples of production code developers could reasonably expect to encounter in the wild. Second, to ensure the \textit{internal validity} of our results, we aimed to sample to an extent that would enable the detailed manual review of each precondition in its code context. Finally, to ensure that our results were \textit{comparable} to an alternative approach, we needed to draw from an existing public dataset. Given these requirements, we leveraged the dataset of Marcilio and Furia~\cite{MF2023a} and subsampled from it to reduce the scale to one that can be manually reviewed. We refer to this dataset as the \textit{WIT dataset}, from the name of the corresponding tool.

The dataset comprises preconditions inferred from methods sampled from two sources: (1) classes from five modules of the Java Development Kit (JDK) version 11, and (2) classes from 46 open source Java projects (OSS). The paper reports two kinds of extracted exception preconditions: \textit{expres}, whose feasibility is fully checked, and \textit{maybes}, which are reported without the same guarantee. We used the 46 OSS project portion of the WIT dataset and excluded the JDK modules because we had used JDK classes as input data to develop the approach. The 46 OSS projects of the WIT dataset thus constitute unseen data for the purpose of our evaluation. Henceforth, references to the WIT dataset are to the OSS portion of the dataset. Excluding the JDK classes, we collected 53\,222 precondition descriptors from the reports in the WIT replication package.

To produce our sample, we applied \tool{} to all Java methods in the WIT dataset. This operation produced 18\,412 precondition descriptors, or only about 35\% of the amount yielded by WIT. From this list, we randomly sampled 150 precondition reports. These reports cover 31 projects, 145 classes, and 150 distinct method $\times$ exception entries. WIT reports a condition for 87 of these entries. We then randomly sampled another 150 precondition descriptors from WIT's 53,222 results. These reports cover 31 projects, 144 classes, and 150 distinct method $\times$ exception entries. Table~\ref{tab:eval-samples} summarizes our evaluation dataset.

\begin{table}[!t]
\centering
\small
\caption{Evaluation Dataset}
\label{tab:eval-samples}
\begin{tabular}{lrrrr}
\toprule
\textbf{Approach} & \textbf{Output} & \textbf{Sample} & \textbf{Projects} & \textbf{Classes}\\
\midrule
\tool{} & 18\,412 & 150 & 31 & 145\\
WIT & 53\,222 & 150 & 31 & 144\\
\bottomrule
\end{tabular}
\end{table}

The two main parameters in this sampling approach are (1) the decision to sample per precondition descriptor as opposed to per method, and (2) the decision to sample 150 precondition descriptors per approach. The justification for our choice of precondition descriptor as the unit of analysis is simply that in operational conditions these are the data items developers would be inspecting. The justification for our selection of 150 items per approach is that we budgeted 25 hours for the focused manual review of results (300 items $\times$ an estimated average of five minutes per item). 

Our associated dataset~\cite{replicationPackage} contains the raw \tool{} JSON reports for the 46 projects, the two manually reviewed 150-report samples, and the manual labeling conducted to map the two samples.

\subsection{RQ1: Precision of \tool{}}
\label{s:ResultsPrecision}

We first investigated \textbf{RQ1: What proportion of preconditions reported by \tool{} are correct?}

\paragraph*{Method}

We manually inspected each of the 150 preconditions reported by \tool{} in detail. We checked each reported precondition against the implementation from which it was extracted, tracing through downstream method calls when needed. We identified a precondition report as \textit{correct} only if it satisfies the following requirements: (1) There is a path through the control flow of a method leading to a \code{throw} statement (called an \textit{exception path}); (2) the exception path does not cross the mutation boundary; (3) the reported guard preserves the condition required by the source check, and (4) the variables and expressions in the reported guard refer to valid program entities: method inputs, instance variables accessible to the method, or expressions over them.

\subsubsection*{Results}

In the \tool{} sample, 149 of the 150 reported conditions are correct. This observation corresponds to a sampled precision of 99.3\%, with a Wilson binomial confidence interval of $[96.3\%, 99.9\%]$ at the 95\% level.
The one incorrect report was due to a missing corner case in our implementation. The precondition is in a getter method, and \tool{} transforms the access to a private field into a call to the method containing the precondition, which creates a circular dependency. 

\begin{tcolorbox}[resbox]
\textbf{Answer to RQ1 (Precision)}: 149 of the 150 preconditions reported by \tool{} and randomly-sampled for inspection were correct. The single incorrect result was due to an implementation oversight, as opposed to an inherent limitation of the approach.
\end{tcolorbox}

\subsubsection*{Discussion}
The answer to RQ1 validates our assumption that pattern-based precondition extraction can produce precise output by targeting code patterns that align with developer intent check preconditions in method implementations. However, the inherent trade-off for a high-precision solution is the potential presence of false negative (i.e., missed preconditions). We do not measure recall as part of RQ1. Recall would measure the proportion of all explicit preconditions in the studied methods that \tool{} reports, which requires a ground truth of all such preconditions. Nevertheless, comparison with a path expansion approach (see Section~\ref{s:ResultsComparison}) confirmed that \tool{} did no miss any case identified through this alternative.

\subsection{RQ2: Testability}
\label{s:ResultsTestabilty}

Our answer to RQ1 validated that pattern-based precondition inference, as implemented through \tool{}, identifies preconditions with almost no false positives. The question is then, can we use these reports for software development? We thus sought to answer: \textbf{RQ2: What proportion of correct preconditions reported by \tool{} is testable by client code?}

\paragraph*{Method} As part of our systematic review of the results (see Section~\ref{s:ResultsPrecision}), we also manually assessed to what extent the reported precondition was \textit{testable}. A tool condition is testable if every element (variable or method) needed to evaluate the reported precondition is accessible to client code outside the package. If a precondition check in source code uses hidden state, \tool{} attempts to recover a public query method for the same state as described in Section~\ref{ss:Algorithm}. When such a method is found, \tool{} can report a testable condition. If no expression visible to client code is available, the descriptor remains untestable. For convenience to developers and for the purpose of the research, \tool{} automatically reports whether the reported precondition descriptor is testable.

\paragraph*{Results} Among the 149 correct reports, 125 are testable by client code. In 113 cases, the source condition is already testable and \tool{} preserves that testability. In 12 cases, the condition relies on hidden state, but \tool{} recovers a public expression for the same state. The remaining 24 correct reports depend on inaccessible state and remain untestable.

\begin{tcolorbox}[resbox]
\textbf{Answer to RQ2 (Testability)}: Among the 149 correct reports, 125 are testable by users. \tool{} preserves testability in 113 reports whose source condition is already testable and recovers a testable form in 12 reports whose source condition is inaccessible.
\end{tcolorbox}

\paragraph*{Discussion} Correctness and testability describe different properties of a report. A condition can preserve the required condition from the source check but still depend on state that users cannot observe before calling the method. Most correct \tool{} reports preserve a condition that is already expressed using inputs or public state. The 12 recovered cases show the additional value of variable mapping: the source check uses hidden state, but a public query method provides an equivalent expression that users can test. In the remaining 24 cases, no testable mapping was found, suggesting that these preconditions cannot be easily leveraged for design by contract. This observation motivates improvements in development practices, whereby precondition checks are expressed in terms of code that can be evaluated outside the method. By automatically identifying the testability of precondition checks, \tool{} supports efforts to improve code to make it more suitable for design by contract practices.

\subsection{RQ3: Comparison}
\label{s:ResultsComparison}

We manually reviewed the 150 reports in the WIT sample and compared the results with those from the \tool{} sample. The analysis answers \textbf{RQ3: How does a pattern-based approach compare to a systematic path expansion approach?}

\paragraph*{Method} The samples were drawn independently from the output of each tool. We used the same systematic review process as described in Sections~\ref{s:ResultsPrecision} and \ref{s:ResultsTestabilty}, including the same definitions of \textit{correctness} and \textit{testability}. WIT targets the broader concept of exception behavior, so a report classified as incorrect under our criteria may still correctly describe an exception-triggering condition. The comparison therefore measures the quality of reports that a user may encounter in each output.

\paragraph*{Results}

In the WIT sample, only 57 of the 150 precondition descriptors are correct according to our precondition definition. Most incorrect descriptors are due to mutation boundary violations: the exception-raising statement can be reached after the method has crossed a conservatively-estimated mutation boundary, so the condition is arguably not a clear precondition. Table~\ref{tab:wit-comparison} presents the two independent samples side by side.

\begin{table}[!t]
\centering
\small
\caption{Precision comparison on independent samples from each tool.}
\label{tab:wit-comparison}
\begin{tabular}{lrrrr}
\toprule
Sample & Correct & Incorrect & Precision & 95\% CI \\
\midrule
\tool{} & 149 & 1 & 99.3\% & [96.3, 99.9]\% \\
WIT & 57 & 93 & 38.0\% & [30.6, 46.0]\% \\
\bottomrule
\end{tabular}
\end{table}

The comparison with WIT clearly shows the impact of considering the mutation boundary when extracting preconditions. Among the 93 WIT reports that do not satisfy our precondition definition, 69 describe exception paths that cross the mutation boundary. \tool{} does not report a condition for any of these 69 entries. The mutation boundary therefore removes many exception paths that are valid descriptions of exceptional behavior but are not method-entry preconditions under our definition.

The difference is also visible in the form of the reported guard condition. WIT may inline the code of downstream methods and expose an internal value even when the source condition is already expressed through a public getter. For example, in \code{JavaBeanDescriptor\#getPrimitiveProperty()}, the source check uses \code{isPrimitiveType()}, which is a public method of the same class. \tool{} reports the condition as \code{!(isPrimitiveType())}. WIT reports \code{type != 6}, exposing the internal encoding used by the downstream method instead of the public method accessible to client code. Another case where WIT fails relates to inlining the behavior of lower level library methods. For example, in \code{BitmapCharacterSet\#setStyle(int)}, the source guard checks the size of the receiver's \code{characters} collection. WIT reports the condition as \code{size > 1}, which drops the receiver object expression and retains an unbound variable \code{size}. In contrast, \tool{} reports \code{characters.size() > 1}, preserving the value used by the source.

In terms of overlap, \tool{} reports a condition for 40 of the 57 correct entries in the WIT sample, and all 40 \tool{} reports are correct. The 17 correct WIT reports not covered by \tool{} (i.e., the false negatives) are all explained by our conservative design decisions to (1) only analyze the input program (excluding dependencies), and (2) assume methods have side effect by default. Specifically, some precondition candidates were eliminated because they call methods outside the analyzed program and outside the precondition catalog. In the other cases, dynamically-bound method were conservatively discarded but the manual inspection shows that potential bindings did not result in state mutation. In the \tool{} sample, WIT reports a condition for 87 entries. All 87 \tool{} reports are correct, while 82 of the corresponding WIT reports are also correct under our definition. The remaining five WIT reports do not preserve a correct precondition due to variables that are not in scope at the method entry point.

\begin{tcolorbox}[resbox]
\textbf{Answer to RQ3 (Comparison):} A total of 93 out of 150 (62\%) exception paths reported by the WIT path-expansion approach do not describe likely preconditions because they occur after the method may have mutated the state of the running program and/or involve symbols that are not in scope at the method entry point. Comparison with WIT identified 17 false negatives out of (150 + 17 = 167) precondition reports, all of which due to conservative assumptions in the design of \tool{}.
\end{tcolorbox}

\paragraph*{Discussion} WIT reports feasible conditions that trigger exceptions, including conditions reached after the method has started executing. \tool{} reports explicit checks that act as obligations on the user before execution. The difference in output size follows from the different focus of each approaches. First, the mutation boundary removes exception paths that occur after state changes, including the 69 cases identified in the WIT sample. Second, \tool{} follows checks within the analyzed program and methods listed in its precondition catalog, but does not expand exception behavior from other Java base-library methods. WIT follows such lower-level behavior when constructing exception paths. The smaller output therefore reflects a more targeted analysis scope of \tool{}, rather than a direct measure of recall. The answer to RQ3 thus illustrates how pattern-based precondition extraction is a solution that offers near-perfect precision at the cost of some false negatives in the presence of precondition checks with complex logic or delegation to dynamically-bound and/or externally-implemented methods.

\subsection{RQ4: Value}
\label{s:ResultsValue}

Finally, we sought to better understand the value proposition of our proposed approach. We thus investigated \textbf{RQ4: How does pattern-based precondition extraction complement existing documentation?}

\paragraph*{Method} To evaluate how \tool{} complements documentation, we examined whether the preconditions reported by \tool{} are documented and how much source code analysis is needed to recover them manually. For each method in the \tool{} sample, the first author reviewed its documentation and implementation. We labeled documentation based on the relevant \code{@throws} entry and its description. We separately labeled recovery effort based on the control flow and downstream calls that must be followed to recover the guard from the implementation. Table~\ref{tab:coding-guide} defines both dimensions. \category{None} corresponds to undocumented preconditions, \category{Unexplained} and \category{Unstructured} to fragmentary or unstructured documentation, and \category{Precise} to precise documentation. The quantitative results include only the 149 correct \tool{} reports.

\begin{table*}[!t]
\centering
\small
\caption{Coding Guide for Precondition Documentation}
\label{tab:coding-guide}
\begin{tabularx}{1.0\textwidth}{lXX}
\toprule
\textbf{Code} & \textbf{Description} & \textbf{Example} \\ \midrule
\multicolumn{3}{l}{\textit{Dimension: Documentation (ordinal scale)}} \\ \midrule
\category{None} & The relevant exception or precondition is not documented. & No relevant \code{@throws} entry. \\
\category{Unexplained} & The exception is documented, but the condition is not explained. & \code{@throws IllegalArgumentException} \\ 
\category{Unstructured} & The condition is described in natural language without identifying the exact program values and relations. & \code{@throws IllegalArgumentException if the list is empty.} \\
\category{Precise} & The documentation identifies the relevant program value and the relation that causes the exception. &\code{@throws IllegalArgumentException if index < 0.} \\ \midrule \midrule
\multicolumn{3}{l}{\textit{Dimension: Recovery Effort (ordinal scale)}} \\ \midrule
\category{Direct} & The guard appears directly in the method or in a dedicated validation call. & \code{assert input != null;} \\ \midrule
\category{Intermediate} & Recovering the guard requires a simple substitution, calculation, or helper lookup. & \code{int i = input * 2; assert i > 5;} \\ \midrule
\category{Complex} & Recovering the guard requires combining branches or following nested helper logic. & \code{if (conditionA)...else if (conditionB)...else \{throw Exception\}} \\
\bottomrule
\end{tabularx}
\end{table*}

\paragraph*{Results} Table~\ref{tab:doc-effort-results} reports documentation quality and source recovery effort for the 149 correct \tool{} reports. These labels measure orthogonal dimensions. A precondition can be undocumented but easy to recover if it appears as a direct \code{if} statement followed by \code{throw}, or a downstream validation method. A documented exception can still be hard to understand if the source condition is spread across downstream method calls or control flow.

\begin{table*}[!t]
\centering
\small
\caption{Documentation quality and source recovery effort among correct \tool{} reports.}
\label{tab:doc-effort-results}
\begin{tabular}{lrrrrrrr}
\toprule
\multirow{2}{*}{Reports} & \multicolumn{4}{c}{Documentation} & \multicolumn{3}{c}{Recovery effort} \\
\cmidrule(lr){2-5}\cmidrule(lr){6-8}
& \category{None} & \category{Unexplained} & \category{Unstructured} & \category{Precise} & \category{Direct} & \category{Intermediate} & \category{Complex} \\
\midrule
Correct \tool{} reports & 122 & 6 & 11 & 10 & 108 & 21 & 20 \\
\bottomrule
\vspace{2mm}
\end{tabular}
\end{table*}

The documentation categorization show evidence of poor documentation quality. In 122 of the 149 correct \tool{} reports, the relevant exception or condition is not even documented. In another 17 cases, there is only partial documentation: six cases are \category{Unexplained}, and 11 cases are \category{Unstructured}. Only 10 preconditions have corresponding documentation that names the relevant state and the triggering relation. Overall, 139 of the 149 correct reports are missing precise documentation. 

The categorization of manual recovery effort shows that many preconditions can be retrieved directly via inspection, but not all of them are easy for users to recover. The 108 \category{Direct} cases include direct null checks, simple input checks, assertions, and validation helper methods. The 21 \category{Intermediate} cases require simple code inlining, substitution, or local calculations. The 20 \category{Complex} cases require significantly more effort to extract. Some guards appear only after reconstructing nested \code{if}/\code{else--if}/\code{else} branches, where the exception-throwing branch is reached only after combining earlier branch conditions. Other guards are hidden inside helper methods that also perform setup, input normalization, or queries of the object state. A method call may look like normal setup code, but the callee may perform an additional precondition check before returning. In such cases, a user has to follow the call, mentally inline the relevant guard, and convert it back into input from the client code.

\begin{tcolorbox}[resbox]
\textbf{Answer to RQ4 (Value)} Our sample provides evidence of 122 cases of undocumented and 17 cases of partially-documented preconditions out of 150 inspected. Manually recovering the precondition can require challenging code inspection and reasoning in at least 20 cases.
\end{tcolorbox}

\paragraph*{Discussion} The documentation gap is not limited to preconditions that are difficult to express. Many direct checks are also undocumented, even though their conditions are already present in a form close to what could be shown to users. This suggests that precondition checks and method documentation may be maintained as separate concerns. When checks are placed in helper methods or spread across the control flow, keeping the documentation complete becomes challenging. \tool{} can help connect documentation and code relating to precondition checks. Its descriptors bind the violation signal to the extracted guard and its source check, and express the condition using values available at the method boundary when possible. For direct checks, this information can be collected without requiring users to inspect the implementation. For checks obscured by local calculations, branches, or helper methods, the analysis performs the substitutions needed to recover the condition. These descriptors can therefore provide a basis for completing or checking documentation when the source and its documentation have become inconsistent.

\subsection{Tradeoffs and Threats to Validity}

Our experimental design is centered around a detailed expert inspection of a small sample of results as opposed to a quantitative analysis of large amounts of unvalidated data. This trade-off provides us high confidence that the data items we report on are meaningful, but at the cost that we have not verified how the approach operates on all possible input scenarios in our sampling frame of 46 OSS projects. 

One potential threat to the internal validity comes from the manual labeling required by various components of the evaluation, including of the correctness testability, documentation, and recovery effort associated with precondition descriptors. We design our classification schemes to minimize subjective judgement, but there remains the possibility of human error in any manual inspection process. However, given that each precondition report in the set of 300 had to be reviewed multiple time under different perspectives as part of the research, we estimate the risk of error to be very low and we include our raw data for independent verification. We did not conduct an inter-rater agreement evaluation on the classification because of the high level of effort and expertise involved. In our case, an inter-rater agreement metrics would be more indicative of reviewer motivation and focus during the task than inherent subjectivity in the classification scheme. 

\section{Conclusion}
\label{s:Conclusion}

We presented \tool{}, a static analysis approach for extracting explicit precondition checks from recognized patterns in Java source code. Each precondition descriptor returned by \tool{} records the guard required for a valid call and the violation signal. To the extent possible, the guard expression is transformed into an expression that can be called by client code to check whether the precondition holds. A key aspect of our approach is the computation of a mutation boundary within methods. With this mutation boundary identified, we limit the analysis to checks that occur before the method changes program state. This important functionality allows us to distinguish likely preconditions from general exception-causing paths through the code, reported by previous approaches.

Our evaluation shows that pattern-based extraction can recover explicit preconditions with high precision. A total of 149 of the 150 sampled reports from our current prototype of \tool{} were correct. For several preconditions originally expressed in terms of hidden state, our approach was able to recover the equivalent publicly-testable checks. The remaining untestable conditions revealed contract design issues in which client code must satisfy a precondition without access to the relevant state information. The comparison with an existing path expansion approach also confirmed that our novel approach produces higher precision and more useful results, at the cost of some potential false negatives caused by a conservative approach to estimating state mutation. Finally, inspection of the documentation of sampled methods showed that the preconditions we extract were seldomly documented at the same level of detail as the code checks. Manually recovering these preconditions is nontrivial, especially when the recovery requires inspecting nested method calls. 

\tool{} shows that explicit precondition checks contain contract information that is often missing from documentation or hard to understand. Its output can support documentation and can also identify untestable preconditions that may require changes to the method implementation. Future work can improve the analysis of collections, loops, and dynamic dispatch to improve recall while preserving the precision of the reported conditions. Another direction is to use \tool{} to migrate precondition checks into assert statements, or to inject checks in client code, thus helping developers validate preconditions before a method call.

\section*{Acknowledgements}
This work is funded by the Natural Sciences and Engineering Research Council of Canada (NSERC). The authors are grateful to Bhagya Chembakottu and Edwin Kim for feedback on the paper. 

\bibliographystyle{IEEEtran}
\bibliography{references}

\end{document}